# A Novel Framework for the Automated Characterization of Gram-Stained Blood Culture Slides Using a Large-Scale Vision Transformer

Jack McMahon[1], Naofumi Tomita[2], Elizabeth S. Tatishev[1], Adrienne A. Workman[3], Cristina R Costales[4], Niaz Banaei[4,5,6], Isabella W. Martin[3], Saeed Hassanpour[1,2,7*]

[1] Department of Computer Science, Dartmouth College, Hanover, NH 03755, USA
[2] Department of Biomedical Data Science, Geisel School of Medicine at Dartmouth, Hanover, NH 03755, USA
[3] Department of Pathology and Laboratory Medicine, Dartmouth-Hitchcock Medical Center, Lebanon, NH 03756, USA
[4] Department of Pathology, Stanford University School of Medicine, Stanford, CA 94305, USA
[5] Department of Medicine, Division of Infectious Diseases and Geographic Medicine, Stanford University School of Medicine, Stanford, CA 94305, USA
[6] Clinical Microbiology Laboratory, Stanford Health Care, Palo Alto, CA 94304, USA
[7] Department of Epidemiology, Geisel School of Medicine at Dartmouth, Hanover, NH 03755, USA

## ABSTRACT

This study introduces a new framework for the artificial intelligence-assisted characterization of Gram-stained whole-slide images (WSIs). As a test for the diagnosis of bloodstream infections, Gram stains provide critical early data to inform patient treatment. Rapid and reliable analysis of Gram stains has been shown to be positively associated with better clinical outcomes, underscoring the need for improved tools to automate Gram stain analysis. In this work, we developed a novel transformer-based model for Gram-stained WSI classification, which is more scalable to large datasets than previous convolutional neural network (CNN) -based methods as it does not require patch-level manual annotations. We also introduce a large Gram stain dataset from Dartmouth-Hitchcock Medical Center (Lebanon, New Hampshire, USA) to evaluate our model, exploring the classification of five major categories of Gram-stained WSIs: Gram-positive cocci in clusters, Gram-positive cocci in pairs/chains, Gram-positive rods, Gram-negative rods, and slides with no bacteria. Our model achieves a classification accuracy of 0.858 (95% CI: 0.805, 0.905) and an AUC of 0.952 (95% CI: 0.922, 0.976) using five-fold nested cross-validation on our 475-slide dataset, demonstrating the potential of large-scale transformer models for Gram stain classification. We further demonstrate the generalizability of our trained model, which achieves strong performance on external datasets without additional fine-tuning.

# INTRODUCTION

A Gram stain is a routine diagnostic test performed to help preliminarily identify the causative agent of infection. Gram stains provide crucial early data points for guiding the proper course of antimicrobial therapy to treat infections (1, 2). This study specifically focuses on the diagnosis of bloodstream infections (BSI), a type of infection caused by the presence of bacteria in a patient's blood which can lead to sepsis and be life-threatening. Previous studies report that quick identification of pathogens involved in a BSI can be critical to the success of patient treatment (3). BSIs can have in-hospital mortality rates above 20%, and the correct choice of antimicrobial agent is key for the treatment of patients with BSI (4). In recent years, rapid molecular diagnostics assays, such as the Blood Culture Identification 2 (BCID2; BioMérieux, Marcy-l-Etoile, France), the cobas ePlex Blood Culture Identification Panel (Roche, Basel, Switzerland) and Verigene (DiaSorin, Saluggia, Italy) assays, have advanced the early detection of causative pathogens in BSIs, providing species-level identification as well as detecting key markers of antimicrobial resistance in samples from positive blood cultures (5). However, these assays must be paired with Gram stain results, either to determine whether to run a Gram-positive or Gram-negative panel or to corroborate the validity of the molecular result before reporting, depending on the assay.

Gram stain analysis remains a manual and time-consuming process whereby medical laboratory scientists analyze stained slides under a microscope to interpret the morphology of any microorganisms that may be present (6). Manual slide analysis consumes valuable time for trained medical laboratory scientists and risks error, more often in cases where visible bacteria are rare or poorly stained. Gram-stain error rates can vary significantly between laboratories, ranging from 0.4% to 2.7%, with discrepancies often involving missed organisms or organisms reported on Gram stain but not recovered in culture (7). Samples can also be falsely flagged for bacterial growth by monitoring systems in 1% to 10% of cases, leading to the preparation of slides with no bacteria that are especially time-consuming to analyze (8). An automated solution for Gram stain characterization can free up valuable time in clinical microbiology laboratories and enhance the efficient use of rapid molecular diagnostic assays by reducing labor demands and streamlining workflow. Additional benefits of rapid Gram stain characterization paired with molecular diagnostics include the ability to use a more targeted narrow-spectrum antimicrobial agent which can be less harmful to beneficial microbes in the body and may mitigate the development of antibiotic-resistant bacteria (9, 10).

Digital microscopy has been successfully combined with deep learning methods to automate slide analysis in other applications. Digital microscopy involves digitizing microscope slides into whole-slide images (WSIs), which provide a tiled view of the slide at various resolutions. WSIs typically have large file sizes, often reaching several gigabytes or more (**Figure 1**). WSI analysis has been extensively explored in computational pathology, particularly for cancer diagnosis, with numerous clinical tools and specialized pipelines developed for processing WSIs as inputs to machine learning models (11, 12). Several factors have made such progress more difficult for Gram stain interpretation. Scanning microbiology slides introduces challenges that are less of an issue when scanning solid tissue, such as uneven fields of focus and background staining leading to blurry regions on WSIs (13). Residual oil on slides from clinical examination can also cause difficulties for digitization. These issues and the variable dispersion and staining of background materials and bacteria make it difficult to accurately segment Gram stain WSIs and process them in the same way as in histology slides.

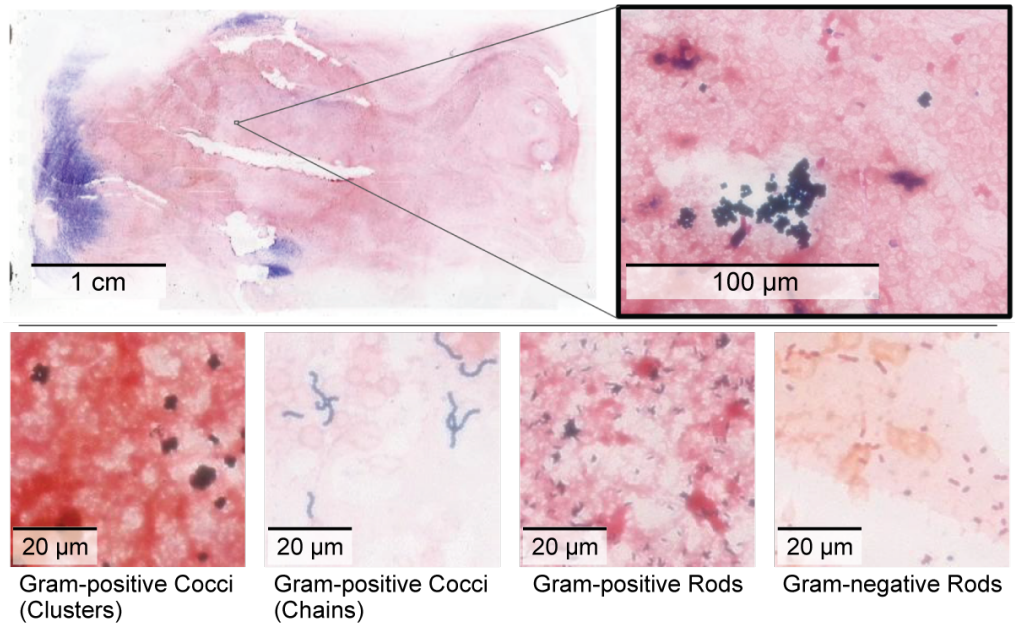

**Figure 1.** Sample images from blood cultures in the Dartmouth-Hitchcock dataset, digitized at 40x magnification. **Top:** A scale comparison between a WSI scan and Gram-positive cocci in clusters. **Bottom** Representative patches depicting four categories of bacterial morphologies.

Despite these challenges, previous studies have shown that deep learning methods can be effective for bacterial classification on datasets of cropped bacterial image patches (typically resized to 224x224 pixel dimensions or similar), distinguishing both Gram-positive/negative status and clinically relevant cellular morphologies (6, 14–16). In 2018, Smith et al. introduced one of the first proof-of-concept studies for Gram-stained WSI classification. They used a convolutional neural network (CNN) -based model that was trained on manually annotated 146x146 pixel image patches from 189 blood culture slides and aggregated predictions from patches to obtain a WSI-level label. Using this approach, they demonstrated a classification accuracy of 92.5% while predicting slides across four separate classes, with the result excluding slides misclassified as background. This work also developed the concept of using deep learning models to extract diagnostically relevant patches for microbiologist review (13). In 2021, Alhammad et al. used manually labelled patches to train a CNN to identify and remove background areas prior to further analysis of Gram-stained WSIs, and in 2024 Walter et al. developed a CNN-based model as a clinical tool for the analysis of Gram-stained WSIs (8, 17). The model was evaluated by trained microbiologists and, rather than making a WSI-level prediction, it was designed to identify and classify diagnostically relevant image patches for review by microbiologists.

In this work, we propose GramViT, a novel vision transformer approach for automating Gram stain analysis, building upon recent advances in computer vision and digital pathology. The Transformer architecture has underpinned some of the most important recent advances in deep learning, such as ChatGPT, using self-attention to model complex relationships in sequences of input data (18). Vision Transformers have been used successfully to improve model performance and remove the need for manual patch-level annotations in computational pathology (19–21). Self-attention allows Vision Transformers to learn to identify diagnostically relevant slide regions in a self-supervised manner, enabling the creation of unprecedentedly large foundation models for computational pathology (22, 23). Recently, LongNet and LongViT have been introduced as an effective approach to train on Gigapixel size images using dilated attention (24, 25). GramViT lays out a framework to apply these techniques to Gram stain analysis, using a pre-trained LongViT model to generate embeddings for large 4,096x4,096 pixel regions, which are randomly sampled during training and systemically aggregated during inference to obtain WSI-level predictions. By this approach, our work aims to bridge the gap toward developing a more robust model for Gram stain analysis that can be trained without manual patch-level annotations and scale efficiently to larger datasets.

## MATERIALS AND METHODS

**Datasets**

This study introduces a newly collected Gram stain dataset from Dartmouth-Hitchcock Medical Center, a tertiary medical center in Lebanon, NH. Between August 2023 and July 2024, a total of 516 Gram-stained blood culture slides obtained during routine patient care were collected, deidentified, assigned a study number, and digitized into WSIs using a Grundium OCUS40 microscope scanner (Grundium, Tampere, Finland) at 40x magnification (0.25 µm/pixel). Slides were prepared from patient samples reported positive for potential bacterial growth by the BD BACTEC™ Blood Culture System (Becton, Dickinson, and Company, Franklin Lakes, NJ). The data collection and usage in our study were approved by the Dartmouth-Health Institutional Review Board (IRB). The study was designed to include the five most common slide types: 1) Gram-positive cocci (GPC) in clusters, 2) GPC in pairs/chains, 3) Gram-positive rods (GPR), 4) Gram-negative rods (GNR), and 5) No bacteria. Other categories, such as Gram-negative cocci, yeast, or slides with multiple morphologies, were not included due to the rarity of these types in the DHMC's cohort. After a quality assurance review, 26 scanned slides were excluded from the study as they fell into rare categories rather than the five major categories considered in this study. Additionally, 15 other scanned slides were excluded due to digitization or staining artifacts, or poor image focus quality. The remaining 475 Gram-stained slides were included in the study, as shown in **Table 1**. Of the 475 included slides, 11 were initially reported incorrectly by laboratory technicians and later corrected, representing an error rate of 2.32%.

| Bacterial Subgroup | WSI Count |
| --- | --- |
| Gram-positive Cocci in Clusters | 184 |
| Gram-positive Cocci in Pairs/Chains | 68 |
| Gram-positive Rods | 37 |
| Gram-negative Rods | 122 |
| No Bacteria | 64 |
| Total | 475 |

**Table 1.** Statistics of the DHMC dataset: WSI counts across selected bacterial subgroups.

To demonstrate the generalizability of our approach, we utilized two external datasets for additional validation (**Table 2**). The first external dataset consists of medium-sized cropped Gram stain images provided by collaborators at Stanford Health. This dataset contains 1 to 3 large images per slide, scanned at 80x magnification using MoticEasyScan Infinity Scanner (Motic, Hong Kong), collected from a total of 32 slides. These include blood culture infections and samples from various other infection sites, such as wounds and cerebrospinal fluid. Due to the limited dataset size and the lack of detailed labels for many slides, we focused on binary classification between Gram-positive and Gram-negative bacteria. After excluding slides with multiple or no bacterial labels, 27 slides remained in the Stanford dataset. The second external dataset is a publicly available test set comprising 1,000 small image crops collected at Medical Faculty Mannheim, Heidelberg University (MHU), from sepsis patients between 2015 and 2019 (11, 26). The MHU dataset contains images classified as either Gram-positive or Gram-negative, however, details regarding the distribution of images per slide, the scan resolution, and the scanner model were not provided.

| Dataset | Classification Scope | Labels | Test Set |
| --- | --- | --- | --- |
| Stanford | Whole-slide level | Gram-positive/Gram-negative | 27 Slides |
| MHU | Patch level | Gram-positive/Gram-negative | 1000 Images |

**Table 2**. Statistics of the external datasets: Stanford and MHU datasets. The Stanford dataset was used to assess WSI-level predictions, while the MHU dataset was used to evaluate patch-level predictions.

**Data Preprocessing**

The OpenSlide library was used for reading and processing the WSIs stored in SVS file formats. A simple image thresholding algorithm was first applied to slides at a lower resolution to identify regions with a color profile consistent with stained material. A sliding window approach was then used to extract non-overlapping 4,096x4,096 pixel regions that contained at least 15% stained material. Optical focus quality can be a significant issue with Gram-stained slides. To address this, regions were further filtered to exclude those with low Laplacian variance, a method that can flag blurry images (26). The number of regions extracted can vary widely depending on the slide, as some slides were almost entirely covered with sample material, while others had large blank spaces. For about 90% of slides, 30 to 750 regions were extracted, with a median of 347. In total, 177,485 regions were extracted from the entire 475-slide dataset (**Figure 2**).

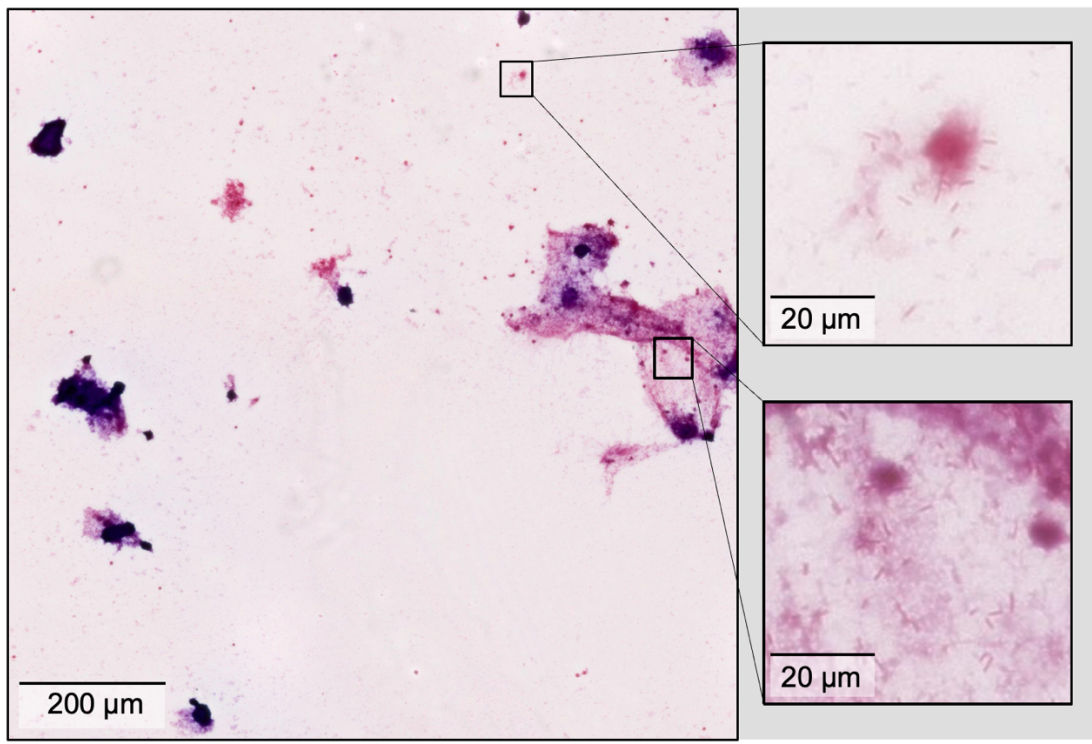

**Figure 2.** A representative 4,096x4,096 region at 40x magnification extracted from a slide with Gram-negative rods in the DHMC dataset. Inset windows show bacterial size and distribution. Unlike previous methods with smaller training regions, GramViT trains on large regions that capture diverse bacterial distributions and background material, improving region-level predictions. Larger region size also increases the likelihood of matching region-level labels to slide-level labels, crucial for effective model training.

**GramViT Training Pipeline**

Our pipeline uses the LongViT vision transformer architecture, which has been pre-trained using The Cancer Genome Atlas (TCGA) dataset to extract histopathological features and generate embeddings for gigapixel-sized histology images (25). Vision transformers, such as LongViT, require large amounts of training data to accurately learn the complex relationships necessary for extracting diagnostically relevant information from whole-slide images. Due to this requirement, it is common to first pre-train a model using self-supervised learning on large datasets to learn how to generate a meaningful embedding. LongViT was pre-trained using the DINO framework for this stage (27). Fine-tuning a pre-trained LongViT model, rather than starting from scratch, allows us to mitigate the challenge posed by the relatively limited size of the DHMC Gram stain dataset. This approach aligns with the transfer learning paradigm, commonly used in other areas of medical image analysis, where pre-trained models are adapted for tasks with limited data availability (28).

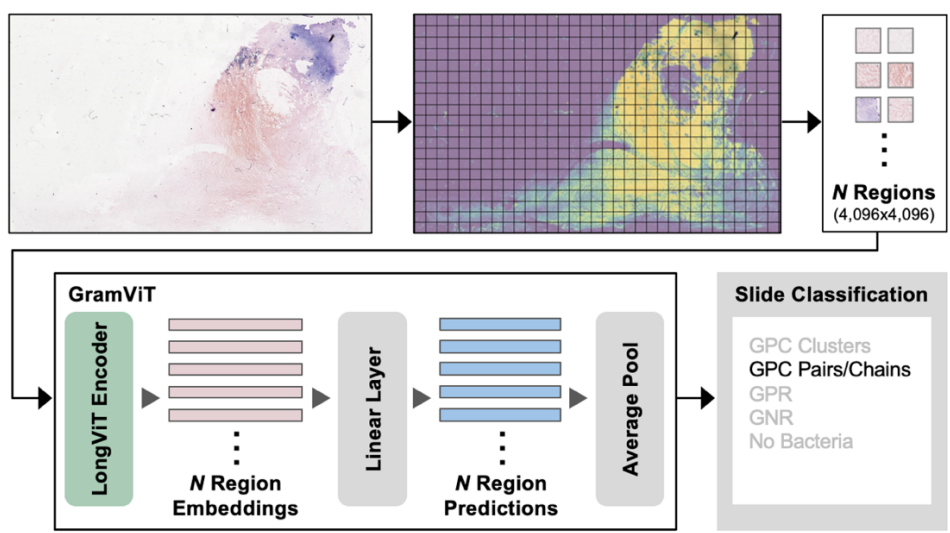

**Figure 3.** Overview of the GramViT training process: A Gram-stained WSI was divided into a grid of non-overlapping regions, with background regions identified and filtered out. A specific number of regions were selected from each sampled WSI to train the LongViT encoder, which produced a 384-dimensional embedding per region. A linear classification layer was then applied to predict class outputs for each embedding, and average pooling of regional class predictions was used to generate a WSI-level classification. During testing, our approach systematically sampled every region of a given WSI for the final inference.

Our training process involved replacing the model's final layer with a linear classification layer that maps the 384-dimensional embedding output of LongViT to the five output classes in this study. During each training epoch, sampling a WSI involved extracting a single 4,096x4,096 pixel region to fine-tune the model (**Figure 3)**. We sample each WSI multiple times per epoch and apply oversampling to minority classes to account for class imbalances during training. Training loss is calculated using cross entropy.

During the testing phase, each region is systematically sampled, and class predictions are average-pooled across regions before calculating performance metrics at the WSI prediction level. GramViT is optimized through our region-sampling approach during training. Relatively large image regions are used as model inputs, rather than downscaling the images during fine-tuning, unlike the approach in the original LongViT paper (25). This is because, in Gram stain analysis, the model relies more on preserving finer details of individual cells rather than capturing larger structures, as is common in cancer histology slide analysis. In our case, downscaling can obscure the distinctions between similar morphologies, such as Gram-positive rods and Gram-positive cocci in pairs/chains. This effect is explored in an ablation study on input image resolution, which is included in the Results section. Additionally, we have access to a relatively small dataset of 475 WSIs in this study. Therefore, using a random region-sampling approach in our training process can be beneficial for increasing the effective training data. Unlike cancer histology slides, Gram-stained blood culture slides originate from a liquid medium, which tends to distribute diagnostically relevant regions more uniformly. As a result, bacteria are typically smeared consistently across visibly stained regions of the slide, and a sufficiently large region chosen at random is likely to contain bacteria corresponding to the slide-level label.

**Evaluation Metrics and Statistical Analysis**

Model predictions are compared to the ground truths for each slide, as listed in the clinical laboratory report. We report model performance using F1-score, accuracy, and area under the receiver operating characteristic curve (AUC), as well as precision, recall, sensitivity, and specificity on a per-class basis. Average metrics are reported using micro-averaging to account for class imbalances and to better reflect performance across the typical composition of slides encountered by microbiology laboratories. The exception is AUC, which is reported using macro-averaging. All metric implementations are sourced from scikit-learn library (Scikit-learn Consortium, Inria Foundation), and 95% confidence intervals (95% CIs) are computed using bootstrapping.

## RESULTS

Experiments were conducted using 5-fold nested cross-validation on the DHMC Gram stain dataset. In each iteration of 5-fold cross-validation, 60% of the data was used for training, with 20% used for validation and 20% used for testing. As discussed above, due to the limited distribution of certain types of bacterial morphologies in the dataset, experiments focused on classification between the five most common Gram-stained slide types (Gram-positive cocci in clusters, Gram-positive cocci in pairs/chains, Gram-positive rods, Gram-negative rods, and no bacteria). Slides that contained yeast, Gram-negative cocci, or multiple labels were excluded from our study.

The experimental setup used 40x magnification and a region size of 4,096x4,096 pixels. Regions were sampled from a grid with no overlap, with one region sampled per WSI during model training. Every region in a WSI was sampled systematically during validation and testing to obtain a stable and comprehensive prediction for each slide. Training used a learning rate of 5e-5 and five warmup epochs. We also used a batch size of 8 and an update frequency of every three steps. Models were trained for 30 epochs, with each WSI sampled up to 11 times per epoch depending on class labels in order to account for class imbalances. Each fold was trained using a Nvidia RTX A6000 GPU (Nvidia, Santa Clara, CA). Testing results aggregate performance on the test set across each of the five folds (**Figure 4** & **Figure 5**).

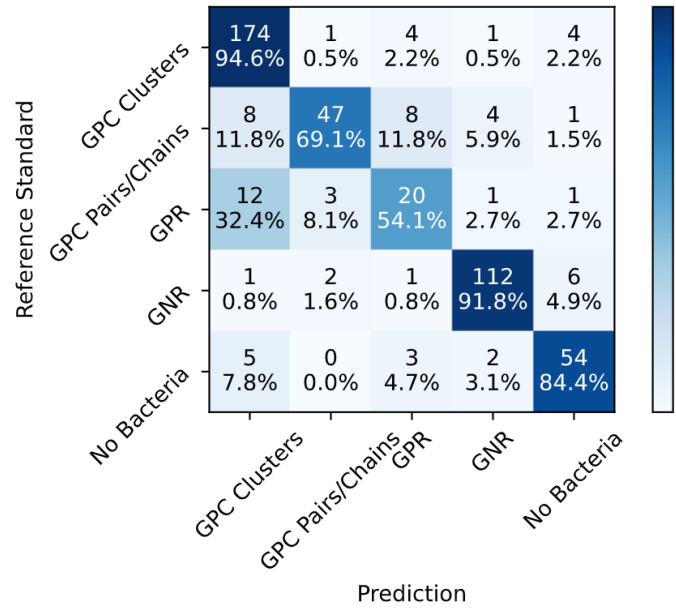

**Figure 4.** GramViT Confusion Matrix with precision/recall breakdown across five bacterial morphologies. Comparing model predictions on the DHMC test set using five-fold nested cross-validation to ground truths determined via pathologist report.

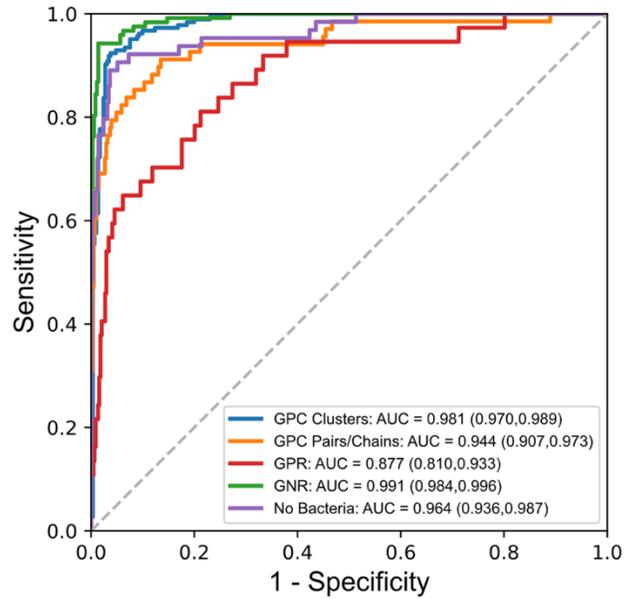

**Figure 5.** GramViT Receiver operating characteristic (ROC) curves and associated area under the curve (AUC) compare the model performance across each of the five classes of WSIs in the DHMC test set.

Overall, GramViT achieved an accuracy of 0.857 (95% CI: 0.810, 0.900) and an AUC of 0.952 (95% CI: 0.922, 0.976). When considering the breakdown of results by category, there is a strong gap in performance between well-represented slide types (i.e., GPC in clusters and GNR), less well-represented slide types (i.e., GPC in Pairs/Chain and No Bacteria), and poorly-represented slide types (i.e., GPR) in the dataset distribution (**Table 3**). Of note, among the 11 slides initially misclassified by laboratory technicians, GramViT correctly predicted the bacterial class in 8.

| Morphology | F1 Score | Precision | Recall/Sensitivity | Specificity |
|---|---|---|---|---|
| **GPC in Clusters** | 0.906 (0.855, 0.948) | 0.871 (0.800, 0.938) | 0.945 (0.892, 0.988) | 0.910 (0.862, 0.958) |
| **GPC in Pairs/Chains** | 0.774 (0.619, 0.880) | 0.888 (0.731, 1.000) | 0.692 (0.500, 0.842) | 0.985 (0.965, 1.000) |
| **GPR** | 0.539 (0.273, 0.739) | 0.555 (0.273, 0.812) | 0.540 (0.278, 0.778) | 0.964 (0.936, 0.989) |
| **GNR** | 0.925 (0.868, 0.971) | 0.934 (0.867, 1.000) | 0.917 (0.833, 0.981) | 0.978 (0.953, 1.000) |
| **No Bacteria** | 0.827 (0.704, 0.926) | 0.816 (0.655, 0.950) | 0.843 (0.696, 0.964) | 0.970 (0.942, 0.994) |
| *Micro-average* | *0.857 (0.810, 0.900)* | *0.857 (0.810, 0.900)* | *0.857 (0.810, 0.900)* | *0.965 (0.951, 0.976)* |

**Table 3**. Results of GramViT predictions by class on the DHMC Gram stain dataset using region resolution at 40x magnification, with 95% confidence intervals.

To benchmark results for GramViT against a CNN-based model, we used the Deepslide framework (12). Deepslide is a sliding-window based framework for microscopy image classification that we trained on the DHMC dataset using patches of size 224x224 pixels without patch-level annotations. The same stain-filtering extraction methodology and five-fold nested cross-validation evaluation were used for both models. We observed that GramViT had higher accuracy, F1, and AUC score than DeepSlide in all metrics (**Table 4**).

| Model | Accuracy | F1 Score | AUC |
|---|---|---|---|
| **DeepSlide** | 0.771 (0.710, 0.820) | 0.771 (0.705, 0.830) | 0.908 (0.880, 0.930) |
| **GramViT** | 0.857 (0.810, 0.900) | 0.857 (0.810, 0.900) | 0.952 (0.922, 0.976) |

**Table 4.** GramViT prediction results benchmarked against a CNN-based model using the Deepslide framework, with 95% confidence intervals.

**Ablation Study**

We conducted an ablation study to validate our choice of region dimensions and image resolution for model training. In total, we trained four models on either 4,096x4,096 or 1,024x1,024 pixel input regions, at 40x magnification or down-sampled to 20x magnification. To balance the amount of training data available to each model per epoch, the 1,024x1,024 pixel models were trained on 16 regions per sampled WSI while the 4,096x4,096 pixel models were trained on only one region per sampled WSI. This evaluation is especially important for Gram stain analysis because it investigates our hypothesis that transformer models will become more effective at bacteria classification and background discrimination when trained on larger input regions. Investigating the effect of different input resolutions aims to determine whether down sampling, a technique that could lead to faster processing and training times, negatively impacts model performance. Our results suggest that down sampling to 20x magnification could, in fact, reduce model performance, though this negative impact is less than that of input region size (**Table 5**).

| Model Input | Accuracy | F1 Score | AUC |
|---|---|---|---|
| 1,024x1,024 at 20x | 0.756 (0.717, 0.794) | 0.756 (0.717, 0.794) | 0.899 (0.871, 0.921) |
| 1,024x1,024 at 40x | 0.787 (0.751, 0.824) | 0.787 (0.751, 0.824) | 0.916 (0.891, 0.937) |
| 4,096x4,096 at 20x | 0.793 (0.757, 0.83) | 0.793 (0.757, 0.83) | 0.927 (0.905, 0.945) |
| 4,096x4,096 at 40x | 0.865 (0.835, 0.896) | 0.865 (0.835, 0.896) | 0.964 (0.948, 0.976) |

**Table 5.** Ablation study comparing the impact of input region size and resolution on GramViT model performance. Models were trained for 30 epochs using five-fold cross-validation and evaluated for performance on the validation sets, totaling 475 slides across all folds.

**Evaluation on External Dataset**

To ensure that GramViT is robust in the presence of variability in bacterial morphology and staining techniques and generalizes beyond the DHMC dataset, we validated the trained model on two external datasets (**Table 2**). Extracted regions from the Stanford dataset were down-sampled to 40x resolution. In cases where the Stanford dataset contained multiple crops from a single slide, we aggregated regions across all crops to make a single slide-level prediction in a manner consistent with that used on the DHMC dataset. Since the MHU dataset contains small image crops no slide-level information is provided, each crop was characterized independently. This is consistent with the approach used in the original MHU study. For both datasets, GramViT was trained on the DHMC dataset and then applied directly to characterize slides without any fine-tuning (**Table 6**).

| Dataset - Method | Accuracy | F1 Score | AUC |
| --- | --- | --- | --- |
| **Stanford - Deepslide** | 0.702 (0.640, 0.760) | 0.702 (0.640, 0.760) | 0.675 (0.366, 0.963) |
| **Stanford - GramViT** | 0.926 (0.885, 0.960) | 0.926 (0.885, 0.960) | 0.865 (0.634, 0.992) |
| **MHU - PoolFormer** (29) | 0.951* | - | - |
| **MHU - Deepslide** | 0.528 (0.455, 0.600) | 0.528 (0.455, 0.600) | 0.528 (0.505, 0.552) |
| **MHU - GramViT** | 0.898 (0.855, 0.935) | 0.898 (0.855, 0.935) | 0.951 (0.948, 0.954) |

**Table 6.** Results compare models trained on the DHMC dataset and applied for binary classification between Gram-positive and Gram-negative bacteria on each external dataset. The PoolFormer model was trained on the MHU dataset, and its result is sourced from the original MHU study*.

We also compared the performance of GramViT with a CNN-based Deepslide model for external dataset evaluation. Both models were trained on the DHMC and evaluated for binary classification on external datasets without fine-tuning. GramViT demonstrated significantly better generalization on the Stanford dataset, achieving an AUC of 0.8651 (95% CI: 0.6337, 0.9917), compared to Deepslide's AUC of 0.675 (95% CI: 0.366, 0.963). A similar trend was observed with the MHU dataset, where GramViT achieved an AUC of 0.9507 (95% CI: 0.9477, 0.9539) versus Deepslide's AUC of 0.528 (95% CI: 0.505, 0.552). However, neither model matched the accuracy reported with the PoolFormer model in the original paper (29).

## DISCUSSION

This study proposes GramViT, a framework that applies vision transformer-based methodologies to the classification of Gram-stained WSIs. GramViT aims to provide an effective solution for training on Gram-stained WSIs in a weakly supervised manner, eliminating the burden of obtaining manual patch-level data annotations, which is a major bottleneck in scaling up model training for larger datasets. Our results, an accuracy of 0.857 (95% CI: 0.810, 0.900) and an AUC of 0.952 (95% CI: 0.922, 0.976), demonstrate the model's ability to accurately characterize and classify Gram-stained WSIs. Notably, GramViT successfully identified 8 out of the 11 slides from the DHMC dataset that had been misclassified in initial laboratory reports, highlighting its potential to catch details that may be missed by laboratory technicians when analyzing slides.

This work advances prior methods by incorporating significantly larger input regions for weakly supervised model training. Unlike CNN-based models that typically rely on small 224x224 pixel input patches, GramViT uses 4,096x4,096 pixel regions, which are sufficiently large to reliably encompass bacteria, considering their typical dispersion across smeared slides. By training a LongViT model with these larger input regions, we leverage pretraining for general WSI interpretation across various histopathology datasets, while fine-tuning the embeddings to capture Gram-stain-specific features.

The benefits of GramViT's large input region in our approach compared to CNN-based models are seen in the head-to-head comparison presented in this work. GramViT achieved better characterization of slides within the DHMC dataset, with an accuracy of 0.857 (95% CI: 0.810, 0.900) compared to 0.771 (95% CI: 0.710, 0.820) for the Deepslide CNN-based model. However, the true strengths of our vision transformer-based approach become more evident when evaluating model performance on external datasets. A clinically useful Gram stain classification model must generalize to bacterial features across diverse laboratory and clinical settings. Scans of Gram-stained slides can vary widely in lighting, color profile, and stain characteristics due to different scanners and staining protocols (7). GramViT excels in binary classification without fine-tuning, achieving an accuracy of 0.926 (95% CI: 0.885, 0.960) on the Stanford dataset and 0.898 (95% CI: 0.855, 0.935) on the MHU dataset. In contrast, the CNN model struggles to perform, with accuracies of 0.702 (95% CI: 0.640, 0.760) and 0.528 (95% CI: 0.455, 0.600) on the Stanford and MHU datasets, respectively. This highlights the vision transformer's ability to identify robust and transferable features essential for bacterial characterization, regardless of laboratory-specific staining and scanning variations.

While GramViT's performance on the MHU dataset is strong, with an accuracy of 0.898 (95% CI: 0.855, 0.935), the PoolFormer model reported in the original paper still outperforms it with an accuracy of 0.951. It is important to note that the PoolFormer model was trained specifically on the MHU training set for small-crop classification, whereas GramViT, trained on the DHMC dataset for WSI-level classification, was applied out-of-the-box without prior exposure to small image crops. Although fine-tuning on the MHU dataset could improve GramViT's performance, these strong out-of-the-box results are encouraging. Any practical implementation of automated characterization tools for Gram stain analysis will need to perform well in new hospital settings without requiring extensive fine-tuning for each specific setup.

Furthermore, our study provides evidence that Gram stain interpretation benefits from using larger region inputs at higher resolution. Our ablation study demonstrates that the model trained on 4,096x4,096 pixel regions at 40x magnification outperforms models trained with smaller 1,024x1,024 pixel regions or 20x magnification. This suggests that further improvements may be achieved by either increasing the resolution beyond 40x magnification or using even larger input regions.

Common applications of transformers in computational pathology often utilize datasets containing 2,000 to 10,000 slides or more, indicating that larger dataset sizes could yield substantial benefits (21, 25). This is further supported by our findings, where the model's performance was notably better for bacterial classes that were more prevalent in the dataset. Despite oversampling minority classes during training to address class imbalance, Gram-positive cocci in clusters and Gram-negative rods were identified far more accurately than other bacterial types. For example, Gram-positive cocci in clusters, with 184 WSIs, had a sensitivity (recall) of 0.945 (95% CI: 0.892, 0.988), whereas Gram-positive rods, with only 37 WSIs, had a sensitivity (recall) of 0.540 (95% CI: 0.278, 0.778). This suggests that increasing the amount of data, particularly for less common bacterial morphologies, could improve the model's ability to learn and distinguish these classes more effectively.

Our study's limitations include the dataset size and the limited diversity of bacterial morphologies. Although the included morphologies represent some of the most common Gram-stained slide types, other clinically relevant bacteria, such as Gram-negative cocci, slides with mixed bacterial morphologies, and non-bacterial pathogens like yeast, were not included. Additionally, the lack of large-scale, publicly available Gram stain datasets restricted further exploration of model generalizability. To address these limitations, future work will focus on expanding data collection across multiple sites and phases, as well as closer collaboration with clinical practitioners. This will facilitate the transition of GramViT from a proof-of-concept model to a clinical laboratory tool. A key aspect of this transition will involve developing a visualization platform that leverages vision transformer attention weights to highlight critical regions of slides for clinical review.

Given the clinical significance of BSIs and the crucial role of Gram stains in informing early treatment decisions, this work establishes a robust framework for developing a practical tool to assist clinical microbiology laboratories in characterizing Gram-stained bacteria more quickly and reliably. Our findings demonstrate that GramViT is a scalable model capable of handling expanding datasets without the need for time-consuming patch-level annotations, while effectively learning features that generalize across Gram-stain images from different laboratories. The potential clinical impact of this work is substantial, as it could streamline diagnostic workflows, reduce diagnostic delays, and enhance the accuracy of early infection management, ultimately improving patient outcomes in critical care settings. By providing a more efficient and reliable approach to bacterial classification, GramViT has the potential to significantly advance the standard of care in microbiology laboratories.

## ACKNOWLEDGMENTS

We thank Megan McKenzie for her assistance with the slide digitization for the DHMC Gram stain dataset.

## CODE AVAILABILITY

The code for GramViT is available at: https://github.com/BMIRDS/GramViT.